\documentclass{article}
\usepackage{epsfig}
\usepackage[caption=false]{subfig}

\usepackage[final]{neurips_2021}

\usepackage[utf8]{inputenc} 
\usepackage[T1]{fontenc}    
\usepackage{hyperref}       
\usepackage{url}            
\usepackage{booktabs}       
\usepackage{amsfonts}       
\usepackage{nicefrac}       
\usepackage{microtype}      
\usepackage{xcolor}         

\title{Reynolds Stress Modeling Using Data Driven Machine Learning Algorithms}

\author{%
J. P. Panda\\
    Assistant Professor,\\
  DIT University,\\ 
  Dehradun, UK, India\\
  Email: jppanda.iit@gmail.com
}

\begin{document}

\maketitle

\begin{abstract}
Fluid turbulence is an important problem for physics and engineering. Turbulence modeling deals with the development of simplified models that can act as surrogates for representing the effects of turbulence on flow evolution. Such models correspond to a range of different fidelities, from simple eddy-viscosity based closures to Reynolds Stress Models. Till now the focus of the data driven turbulence modeling efforts has focused on Machine Learning augmented eddy-viscosity models. In this communication we illustrate the manner in which the eddy-viscosity framework delimits the efficacy and performance of Machine learning algorithms. Based on this foundation we carry out the first application of Machine learning algorithms for developing improved Reynolds Stress Modeling based closures for turbulence. Different machine learning approaches are assessed for modeling the pressure strain correlation in turbulence, a longstanding problem of singular importance. We evaluate the performance of these algorithms in the learning dataset, as well as their ability to generalize to different flow cases where the inherent physical processes may vary. This explores the assertion that ML-based data-driven turbulence models can overcome the modeling limitations associated with the traditional turbulence models and ML models trained with large amounts of data with different classes of flows can predict flow field with reasonable accuracy for unknown flows with similar flow physics.   
\end{abstract}

\section{Introduction}
Fluid turbulence represents one of the most important unsolved problems in classical physics and is important to different disciples of engineering. Engineering studies for turbulent flows, such as those found over aircraft, utilize Computational Fluid Dynamics (CFD) based simulations that utilize turbulence models to account for the effect of turbulence on flow evolution. Such physics based turbulence models have many shortcomings and recent research has focused on using Machine Learning algorithms to develop data driven turbulence models. Such data driven approaches can be applied for development of data-driven turbulence modeling \cite{duraisamy2019turbulence}, where, the problem is posed as one of supervised learning, where the model attempts to  minimize the prediction error over a training data-set. This data required for training, validation and testing the ML models can be obtained from experimental investigations, high fidelity DNS and LES data-sets of turbulent flows.

Such efforts of augmenting classical physics based turbulence modeling efforts with data driven machine learning approaches have gained popularity over the recent past, and various investigators have exhibited successes with such applications. \cite{singh2017machine} developed model augmentations for Spalart-Allmaras(SA) turbulence model using adjoint based full field inference using experimentally measured lift coefficient data. These models forms are reconstructed using neural networks, and applied in CFD solver, to predict flow in different operating conditions. \cite{tracey2015machine} used a shallow neural network (one hidden layer) to model the source terms of the SA turbulence model. \cite{parish2016paradigm} learned a turbulence production term using machine learning and applied that to the $k$-equation of the $k-\omega$ turbulence model. \cite{maulik2018data} used neural networks to model eddy viscosities for RANS simulations. \cite{ling2016reynolds} used a tensor basis neural network to model the Reynolds stress anisotropy. \cite{zhu2019machine} used  neural networks to construct a mapping function between the turbulent eddy viscosity and the mean flow variables and ML model completely replaces the partial differential equation model. \cite{weatheritt2016novel} used Gene expression programming to formulate non-linear consecutive stress-strain relationship. The mathematical model was created by using high fidelity and uncertainty measures. The learning method has the capability to produce a constraint free model. \cite{heyse2021estimating} utilized an ensemble learning approach that was constrained by physics based requirements to generate uncertainty bounds for the predictions of RANS models. This is important as the physics constraints were guaranteed to be adhered to, as opposed to other ``physics-inspired" applications where the physics based constraining is a regularization term and need not assure the adherence of the constraint.  \cite{weatheritt2017development} used symbolic regression to model the algebraic form of the Reynolds stress anisotropy tensor. The equations were trained using hybrid RANS/LES data and the new model was employed in RANS closure to test the prediction of model in 3D geometries. \cite{schmelzer2020discovery} discovered algebraic Reynolds stress models using sparse regression and they have used high fidelity LES/DNS data for training and cross validation of the model. There case of separated flows were considered, those are periodic hills, converging-diverging channel and curved backward facing step. The prediction of the machine learnt model was better than the $k-omega$ SST model. \cite{huijing2021data} used the model developed by \cite{schmelzer2020discovery} to predict the fully three dimensional high Reynolds number flows, e.g. wall mounted cubes and cuboids. \cite{fang2020neural} used neural networks to model the Reynolds stress anisotropy using neural networks. They proposed different modification of the neural network structure to accommodate effect of Reynolds number, non-locality and wall effects into the modeling basis. With such distinct feature injection, significant improvement of model prediction was observed. \cite{beck2019deep} proposed a novel data-driven strategy for turbulence modeling for LES using artificial neural networks.

While there have been many investigations toward augmenting RANS modeling using machine learning, their focus has been on simplified eddy-viscosity based models. There has been little research to extend the potential of  Reynolds Stress Modeling approach by utilizing machine learning algorithms \cite{panda2021modelling}. This is a central novelty of this investigation. In the Reynolds Stress Modeling approach separate models are formulated for the terms in the Reynolds Stress Transport Equation, where each such term represents a different turbulence transport process. These transport processes include turbulent diffusion, rotational effects, rate of dissipation and the pressure strain correlation \cite{panda2018representation}. While high fidelity models for all these terms are important, accurate and robust modeling of the pressure strain correlation term is a long standing challenge in turbulence modeling \cite{panda2019review}. The pressure strain correlation term represents physics responsible for the transfer of energy between different components of the Reynolds stress tensor \cite{mishra2010pressure}. It is responsible for the non-local interactions in turbulent flows, the initiation of instabilities in rotation dominated flows, the return to isotropy observed in decaying flows, etc \cite{mishra2013intercomponent}. While classical models have been developed for the pressure strain correlation term, such physics driven models have many limitations in their ability to account for streamline curvature effects, realizability requirements, their performance in complex engineering flows \cite{mishra2017toward}. In this work we have modeled the pressure strain correlation of turbulence using three different machine learning approaches, those are Neural Networks, Random Forests and Gradient Boosted decision trees. The data driven models developed with these algorithms are trained and tested for DNS data of turbulent channel flow at different Reynolds numbers. We have grouped the data-sets in 4 different combination to perform 4 distinct training and testing of the ML models. Bayesian Optimization (BO) was utilized to find the best hyper-parameters of the ML models. One main advantage of hyper-parameter optimization is that it reduces the chance of over-fitting in tree based algorithms and also enhances the generalizability of the model.

\begin{table}
\begin{center}
 \begin{tabular}
 {||c  c c||} 
 \hline
 Case & Training Set & Testing Set \\ [0.5ex] 
 \hline\hline
 1 & $Re_\lambda=550,1000,2000$  & $Re_\lambda=5200$ \\ 
 \hline
 2 & $Re_\lambda=550,1000,5200$ &  $Re_\lambda=2000$ \\
 \hline
 3 & $Re_\lambda=550,2000,5200$ &  $Re_\lambda=1000$ \\
 
 \hline
 4 & $Re_\lambda=1000,2000,5200$ & $Re_\lambda=550$ \\ [1ex] 
\hline
\end{tabular}
\end{center}
\caption{Four training and test cases for the turbulent channel flow\label{t2}}
\end{table}
\section{Data driven turbulence modeling with machine learning}
\label{sec: Data driven turbulence}
The dataset used in the modeling of the pressure strain correlation of turbulence was obtained from the DNS simulations of \cite{lee2015direct}. We have used four distinct data-sets for training and testing of the ML models. The four training and testing data-sets are presented in table\ref{t2}.

The functional mapping for the ML model for the pressure strain correlation can be written as: 
\begin{equation}
\phi_{uv}=f_1(b_{uv}, \epsilon, \frac{du}{dy}, k)
\end{equation}
We have used the formula: $\alpha^*=\frac{\alpha-\alpha_{min}}{\alpha_{max}-\alpha_{min}}$ for normalizing the inputs to the ML models, so that those will be in the range 0 and 1. 

In this work, we have modeled the pressure strain correlation using three different machine learning algorithms, those are deep neural networks, Random forests and gradient boosted trees. While the ML algorithms are different, the modeling basis (composed of the feature tensors) is the same across model and is identical to that used in classical physics based modeling. This is to ensure that the ML models developed in this investigation are comparable to classical physics based models. In future investigations, we will look into appending to this basis.

\begin{figure*}
\centering
 \subfloat[]{\includegraphics[width=0.5\textwidth]{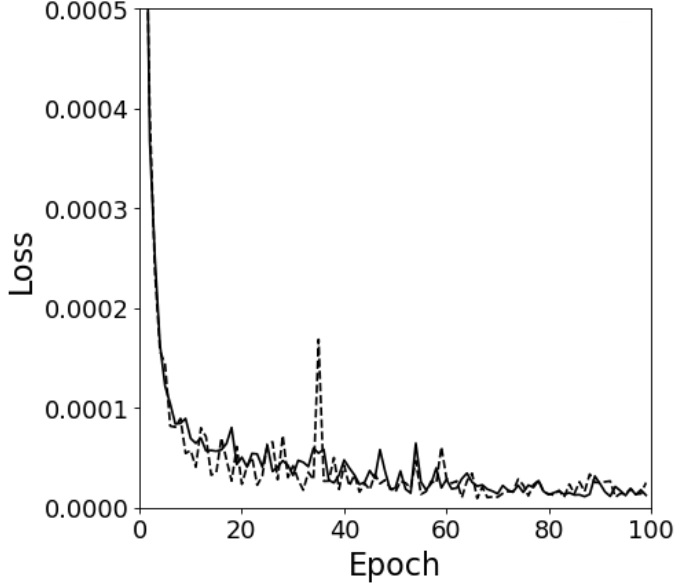}}
 \subfloat[]{\includegraphics[width=0.5\textwidth]{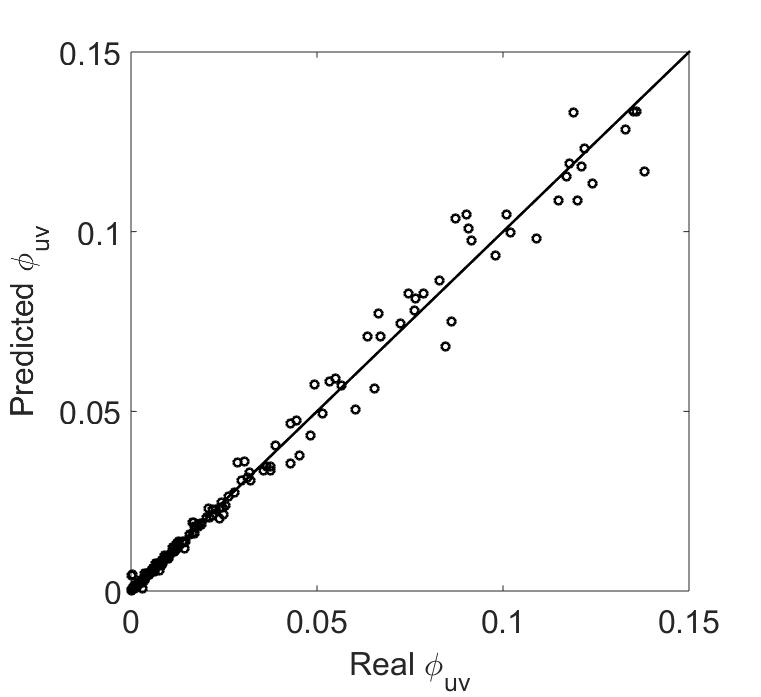}}
 \caption{a) Loss convergence in MLP training for case 4. Solid line and dashed line correspond to training and validation losses respectively, b) DNS simulation samples versus data driven model predictions of the pressure strain correlation for the GBDT training for case 4. \label{fig:5}}
\end{figure*}
\begin{table*}[h]
\begin{center}
\begin{tabular} {c c c c c c c c c} 
  \toprule
  \bfseries  &  \multicolumn{2}{c}{\bfseries Case 1}  &  \multicolumn{2}{c}{\bfseries Case 2} &  \multicolumn{2}{c}{\bfseries Case 3}  &  \multicolumn{2}{c}{\bfseries Case 4} \\
  \cmidrule(lr){1-9}
  \bfseries Models &  \multicolumn{1}{c}{\bfseries Train $R^2$} & \multicolumn{1}{c}{\bfseries Test $R^2$} & \multicolumn{1}{c}{\bfseries Train $R^2$} & \multicolumn{1}{c}{\bfseries Test $R^2$}  &  \multicolumn{1}{c}{\bfseries Train $R^2$} & \multicolumn{1}{c}{\bfseries Test $R^2$} & \multicolumn{1}{c}{\bfseries Train $R^2$} & \multicolumn{1}{c}{\bfseries Test $R^2$}\\
  \midrule
  MLP & 0.967 & 0.978 & 0.978  & 0.984 & 0.921 & 0.891 & 0.992  & 0.892 \\
  RF & 0.971 & 0.978 & 0.949  & 0.966 & 0.962 & 0.988 & 0.984  & 0.885 \\
  GBDT & 0.982  & 0.871  & 0.968  & 0.958 & 0.951 & 0.955 & 0.982  & 0.871 \\
\bottomrule
\end{tabular}
\caption{$R^2$ of $\phi_{uv}^*$ predictions by training-prediction cases}\label{t3}
\end{center}
\end{table*}
\section{Training of the ML models}
\label{sec:training}
All the three ML models are trained for the DNS data of turbulent channel flow for four different test cases. The four different cases are shown in table \ref{t2}. Since the turbulence statistics are available for four different friction Reynolds numbers, in each training and testing phase, we have considered data for three friction Reynolds numbers as training and the other was considered for testing. The loss convergence history of MLP for case 4 is shown in figure.\ref{fig:5} a. In figure.\ref{fig:5} b , the results of DNS simulation samples versus GBDT predictions of the pressure strain correlation are presented. There is very good match between the DNS results and DNS model predictions. There are very few points in the data-set with high values of the pressure strain correlation, this signifies the fact that, the relative inaccuracy of the model predictions for high values of the pressure strain correlation is due to very few such training data samples. 

\begin{figure*}
\centering
\subfloat[]{\includegraphics[width=0.5\textwidth]{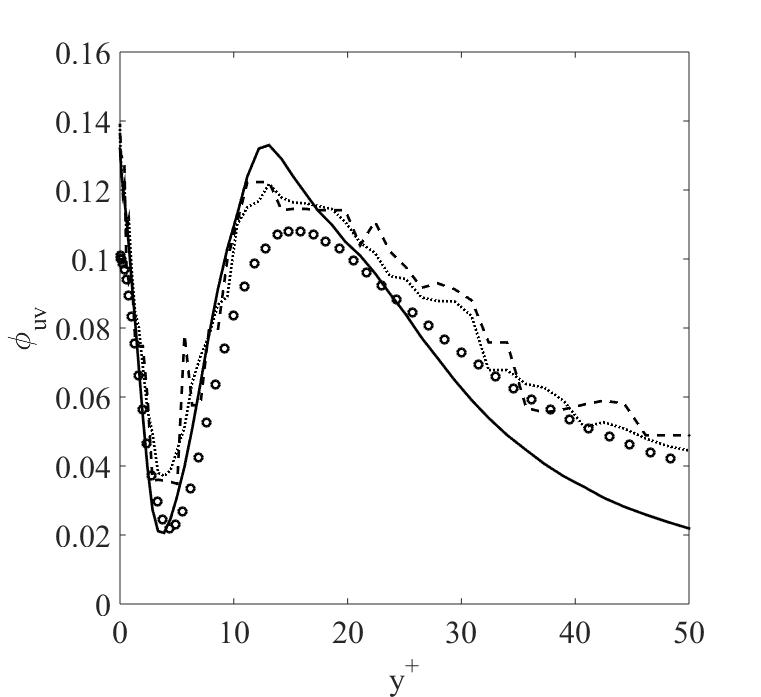}}
\subfloat[]{\includegraphics[width=0.5\textwidth]{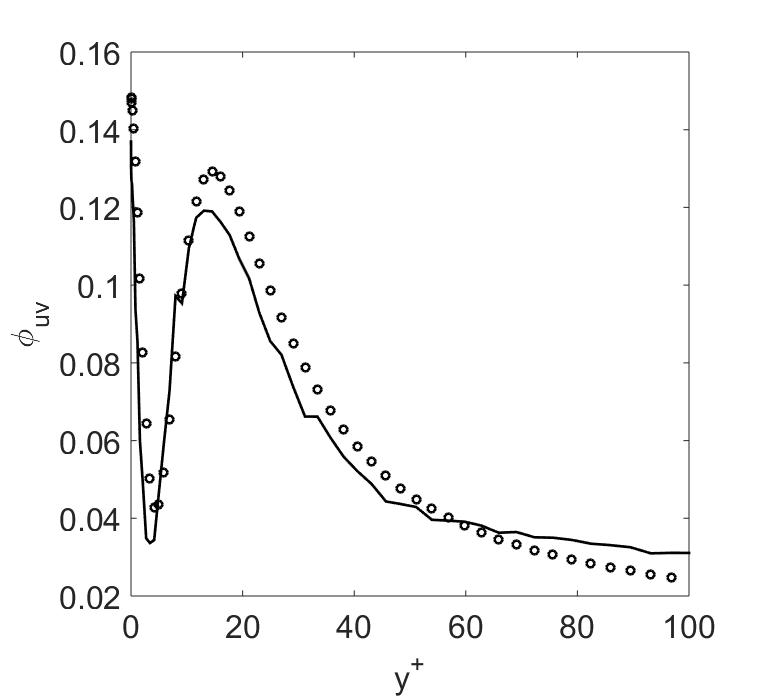}}
 \caption{Prediction of the pressure strain correlation using different machine learning algorithms.Solid line, dashed line and dotted line represent predictions of neural network, gradient boosted trees and random forests respectively. a) case 4, (b) , Prediction of the pressure strain correlation for turbulent Couette flow at $Re_\lambda=550$. \label{fig:7}}
\end{figure*}
\section{Testing of the trained ML models}
\label{sec:testing}
The trained models were tested against testing data for all the four cases as discussed in table.\ref{t2}. The results of testing of the ML models are presented in \ref{fig:7} a. From figure \ref{fig:7} a it is clear that, the GBDT predictions are better in comparison to the predictions of MLP and RF. The neural networks failed to predict the pressure strain correlation for $y+$ values greater than 25 for case 4. The testing results are presented in table \ref{t3}. In most of the testing cases the $R^2$ value was found to be greater than $0.95$, but in few cases like case 4, the $R^2$ value falls below $0.9$ for the testing case. This is because of over-fitting of the models with the training data. The over-fitting problem is common in random forests. The advantage of GBDT over MLP is that the former needs less number of weights in developing the correlation between the inputs and the output. Since the ultimate aim of turbulence modeler is to apply the ML based based into CFD solver, models with less number of weights are always preferred. We have tested the optimized GBDT model against a fully unknown flow e.g. Couette flow in channels. The model predictions are presented in figure.\ref{fig:7} b. It is noticed that the GBDT predictions are matching well the DNS results of \cite{lee2018extreme}.        

\section{Conclusions}
In this article, we discuss rationale for the application of machine learning with high-fidelity turbulence data to develop models at the level of Reynolds stress transport modeling. Then, we ascertain the efficacy of different machine learning algorithms at creating surrogate models for the pressure strain correlation. We have modeled the pressure strain correlation of turbulence using three different machine learning approaches, those are Random forests, gradient boosted tress and artificial neural networks. The input features to the ML models were chosen from the traditional modeling basis of the pressure strain correlation, those are mean strain, turbulence kinetic energy, Reynolds stress anisotropy and turbulence dissipation. The ML models were trained and tested for DNS data of turbulent channel flow at different friction Reynolds numbers. The optimal values of ML model hyper-parameters were optimized using manual search and Bayesian optimization approaches. The feature importances of different input features of the random forest were obtained using mean decrease in impurity method. It was noticed that the mean strain has larger correlation with the pressure strain term. The ML models developed by using data-driven approaches can be utilized in computational fluid dynamics solvers for improved flow predictions in turbulent channel flows.






\bibliographystyle{plainnat}
\bibliography{jafm.bib}

\providecommand{\noopsort}[1]{}\providecommand{\singleletter}[1]{#1}%
\begin{thebibliography}{22}
\providecommand{\natexlab}[1]{#1}
\providecommand{\url}[1]{\texttt{#1}}
\expandafter\ifx\csname urlstyle\endcsname\relax
  \providecommand{\doi}[1]{doi: #1}\else
  \providecommand{\doi}{doi: \begingroup \urlstyle{rm}\Url}\fi

\bibitem[Beck et~al.(2019)Beck, Flad, and Munz]{beck2019deep}
Andrea Beck, David Flad, and Claus-Dieter Munz.
\newblock Deep neural networks for data-driven les closure models.
\newblock \emph{Journal of Computational Physics}, 398:\penalty0 108910, 2019.

\bibitem[Duraisamy et~al.(2019)Duraisamy, Iaccarino, and
  Xiao]{duraisamy2019turbulence}
Karthik Duraisamy, Gianluca Iaccarino, and Heng Xiao.
\newblock Turbulence modeling in the age of data.
\newblock \emph{Annual Review of Fluid Mechanics}, 51:\penalty0 357--377, 2019.

\bibitem[Fang et~al.(2020)Fang, Sondak, Protopapas, and Succi]{fang2020neural}
Rui Fang, David Sondak, Pavlos Protopapas, and Sauro Succi.
\newblock Neural network models for the anisotropic reynolds stress tensor in
  turbulent channel flow.
\newblock \emph{Journal of Turbulence}, 21\penalty0 (9-10):\penalty0 525--543,
  2020.

\bibitem[Heyse et~al.(2021)Heyse, Mishra, and Iaccarino]{heyse2021estimating}
Jan~Felix Heyse, Aashwin~A Mishra, and Gianluca Iaccarino.
\newblock Estimating rans model uncertainty using machine learning.
\newblock \emph{Journal of the Global Power and Propulsion Society. Special
  Issue: Data-Driven Modelling and High-Fidelity Simulations}, pages 1--14,
  2021.

\bibitem[Huijing et~al.(2021)Huijing, Dwight, and Schmelzer]{huijing2021data}
Jasper~P Huijing, Richard~P Dwight, and Martin Schmelzer.
\newblock Data-driven rans closures for three-dimensional flows around bluff
  bodies.
\newblock \emph{Computers \& Fluids}, page 104997, 2021.

\bibitem[Lee and Moser(2015)]{lee2015direct}
Myoungkyu Lee and Robert~D Moser.
\newblock Direct numerical simulation of turbulent channel flow up to $re_tau =
  5200$.
\newblock \emph{Journal of Fluid Mechanics}, 774:\penalty0 395--415, 2015.

\bibitem[Lee and Moser(2018)]{lee2018extreme}
Myoungkyu Lee and Robert~D Moser.
\newblock Extreme-scale motions in turbulent plane couette flows.
\newblock \emph{Journal of Fluid Mechanics}, 842:\penalty0 128--145, 2018.

\bibitem[Ling et~al.(2016)Ling, Kurzawski, and Templeton]{ling2016reynolds}
Julia Ling, Andrew Kurzawski, and Jeremy Templeton.
\newblock Reynolds averaged turbulence modelling using deep neural networks
  with embedded invariance.
\newblock \emph{Journal of Fluid Mechanics}, pages 155--166, 2016.

\bibitem[Maulik et~al.(2018)Maulik, San, Rasheed, and Vedula]{maulik2018data}
Romit Maulik, Omer San, Adil Rasheed, and Prakash Vedula.
\newblock Data-driven deconvolution for large eddy simulations of kraichnan
  turbulence.
\newblock \emph{Physics of Fluids}, 30\penalty0 (12):\penalty0 125109, 2018.

\bibitem[Mishra and Girimaji(2010)]{mishra2010pressure}
Aashwin~A Mishra and Sharath~S Girimaji.
\newblock Pressure--strain correlation modeling: towards achieving consistency
  with rapid distortion theory.
\newblock \emph{Flow, turbulence and combustion}, 85\penalty0 (3-4):\penalty0
  593--619, 2010.

\bibitem[Mishra and Girimaji(2013)]{mishra2013intercomponent}
Aashwin~A Mishra and Sharath~S Girimaji.
\newblock Intercomponent energy transfer in incompressible homogeneous
  turbulence: multi-point physics and amenability to one-point closures.
\newblock \emph{Journal of Fluid Mechanics}, 731:\penalty0 639--681, 2013.

\bibitem[Mishra and Girimaji(2017)]{mishra2017toward}
Aashwin~A Mishra and Sharath~S Girimaji.
\newblock Toward approximating non-local dynamics in single-point
  pressure--strain correlation closures.
\newblock \emph{Journal of Fluid Mechanics}, 811:\penalty0 168--188, 2017.

\bibitem[Panda(2019)]{panda2019review}
JP~Panda.
\newblock A review of pressure strain correlation modeling for reynolds stress
  models.
\newblock \emph{Proceedings of the Institution of Mechanical Engineers, Part C:
  Journal of Mechanical Engineering Science}, page 0954406219893397, 2019.

\bibitem[Panda and Warrior(2018)]{panda2018representation}
JP~Panda and HV~Warrior.
\newblock A representation theory-based model for the rapid pressure strain
  correlation of turbulence.
\newblock \emph{Journal of Fluids Engineering}, 140\penalty0 (8), 2018.

\bibitem[Panda and Warrior(2021)]{panda2021modelling}
JP~Panda and HV~Warrior.
\newblock Modelling the pressure strain correlation in turbulent flows using
  deep neural networks.
\newblock \emph{Proceedings of the Institution of Mechanical Engineers, Part C:
  Journal of Mechanical Engineering Science}, 2021.

\bibitem[Parish and Duraisamy(2016)]{parish2016paradigm}
Eric~J Parish and Karthik Duraisamy.
\newblock A paradigm for data-driven predictive modeling using field inversion
  and machine learning.
\newblock \emph{Journal of Computational Physics}, 305:\penalty0 758--774,
  2016.

\bibitem[Schmelzer et~al.(2020)Schmelzer, Dwight, and
  Cinnella]{schmelzer2020discovery}
Martin Schmelzer, Richard~P Dwight, and Paola Cinnella.
\newblock Discovery of algebraic reynolds-stress models using sparse symbolic
  regression.
\newblock \emph{Flow, Turbulence and Combustion}, 104\penalty0 (2):\penalty0
  579--603, 2020.

\bibitem[Singh et~al.(2017)Singh, Medida, and Duraisamy]{singh2017machine}
Anand~Pratap Singh, Shivaji Medida, and Karthik Duraisamy.
\newblock Machine-learning-augmented predictive modeling of turbulent separated
  flows over airfoils.
\newblock \emph{AIAA journal}, 55\penalty0 (7):\penalty0 2215--2227, 2017.

\bibitem[Tracey et~al.(2015)Tracey, Duraisamy, and Alonso]{tracey2015machine}
Brendan~D Tracey, Karthikeyan Duraisamy, and Juan~J Alonso.
\newblock A machine learning strategy to assist turbulence model development.
\newblock In \emph{53rd AIAA aerospace sciences meeting}, page 1287, 2015.

\bibitem[Weatheritt and Sandberg(2017)]{weatheritt2017development}
J~Weatheritt and RD~Sandberg.
\newblock The development of algebraic stress models using a novel evolutionary
  algorithm.
\newblock \emph{International Journal of Heat and Fluid Flow}, 68:\penalty0
  298--318, 2017.

\bibitem[Weatheritt and Sandberg(2016)]{weatheritt2016novel}
Jack Weatheritt and Richard Sandberg.
\newblock A novel evolutionary algorithm applied to algebraic modifications of
  the rans stress--strain relationship.
\newblock \emph{Journal of Computational Physics}, 325:\penalty0 22--37, 2016.

\bibitem[Zhu et~al.(2019)Zhu, Zhang, Kou, and Liu]{zhu2019machine}
Linyang Zhu, Weiwei Zhang, Jiaqing Kou, and Yilang Liu.
\newblock Machine learning methods for turbulence modeling in subsonic flows
  around airfoils.
\newblock \emph{Physics of Fluids}, 31\penalty0 (1):\penalty0 015105, 2019.

\end{thebibliography}



\end{document}